\documentclass[twocolumn,superscriptaddress,secnumarabic,amssymb, preprintnumbers,nobibnotes, aps, bm,pre,floatfix,longbibliography]{revtex4-2}
\usepackage[english]{babel}
\usepackage[latin9]{inputenc}
\usepackage{graphicx}
\usepackage{multirow}
\usepackage[usenames,dvipsnames]{xcolor}
\usepackage{color}
\usepackage{bm}
\usepackage{mathtools}
\usepackage{amsmath}
\usepackage{rotating}
\usepackage{booktabs}
\newcommand\sbullet[1][.5]{\mathbin{\vcenter{\hbox{\scalebox{#1}{$\bullet$}}}}}
\definecolor{oceanboatblue}{rgb}{0.0, 0.47, 0.75}
\definecolor{orange}{rgb}{1,0.5,0}
\definecolor{goodgreen}{rgb}{0.1,0.5,0}
\definecolor{goodred}{rgb}{0.7,0,0}
\usepackage[colorlinks,urlcolor=goodgreen,citecolor=blue,linkcolor=goodred]{hyperref}
\DeclareMathOperator{\Tr}{Tr}
\usepackage{tikz}
\usepackage{tocvsec2}
\usepackage{scrextend}
\usepackage{lipsum}
\makeatletter
\newcommand{\ee}{\text{e}}
\newcommand{\ii}{\text{i}}

\newcommand{\comment}[1]{}
\usepackage[normalem]{ulem}

\makeatother
\usepackage{babel}

\begin{document}

\title{Tunable Dirac points in a two-dimensional non-symmorphic wallpaper group lattice}
\author{Miguel A. J. Herrera}
\email{ma.jimenez.herrera@gmail.com}
\affiliation{Centro de F\'isica de Materiales (CFM-MPC) Centro Mixto CSIC-UPV/EHU,
20018 Donostia-San Sebasti\'an, Basque Country, Spain}
\affiliation{Donostia International Physics Center, 20018 San Sebasti\'an, Spain}
\author{Dario Bercioux}
\email{dario.bercioux@dipc.org}
\affiliation{Donostia International Physics Center, 20018 San Sebasti\'an, Spain}
\affiliation{IKERBASQUE, Basque Foundation for Science, Euskadi Plaza, 5, 48009 Bilbao, Spain}

\date{\today}

\begin{abstract}
Non-symmorphic symmetries protect Dirac nodal lines and cones in lattice systems. Here, we investigate the spectral properties of a two-dimensional lattice belonging to a non-symmorphic group. Specifically, we look at the herringbone lattice, characterised by two sets of glide symmetries applied in two orthogonal directions. We describe the system using a nearest-neighbour tight-binding model containing horizontal and vertical hopping terms. We find two non-equivalent Dirac cones inside the first Brillouin zone along a high-symmetry path.
We tune these Dirac cones' positions by breaking the lattice symmetries using onsite potentials. These Dirac cones can merge into a semi-Dirac cone or unfold along a high-symmetry path. Finally, we perturb the system by applying a dimerization of the hopping terms. We report a flow of Dirac cones inside the first Brillouin zone describing quasi-hyperbolic curves. We present an implementation in terms of CO atoms placed on the top of a Cu(111) surface.
\end{abstract}
\maketitle
\section*{Introduction}
Since the isolation of single-layer graphene, there has been a growing interest in analysing two-dimensional (2D) systems with low-energy physics described by a Dirac-like electronic dispersion~\cite{Vafek_2014,Goerbig_2017}. In addition to graphene, systems of interest are also 3D topological insulators~\cite{Hasan_2011,Ando_2013,Vafek_2014}, and other 2D materials beyond graphene~\cite{Wang_2015,Galbiati_2019}. The quest for systems hosting Dirac-like features is not only within condensed matter but extended to cold-atoms~\cite{Tarruell_2012} and electronic quantum simulators~\cite{Gomes_2012,gardenier2020p}. Most of these systems share the property of having an underlying crystal structure characterized by a symmorphic space group~\cite{Dresselhaus_2007}. However, there has been an increasing research interest in Dirac-like physics in non-symmorphic crystalline systems~\cite{Dresselhaus_2007,liu2014topological,Young_2015,Shiozaki_2015,Wang_2016,Alexandradinata_2016,Muechler_2016,Kruthoff_2017,wieder2018wallpaper,Ryu_2020}.
A non-symmorphic crystalline system contains a fractional lattice translation combined with either a mirror reflection (glide plane) or a rotation (screw axis). This results in a band-folding with crossings inside the first Brillouin zone (FBZ) boundaries that are protected against hybridization~\cite{schnyder2016nonsymmorphic,Kruthoff_2017,wang2019hourglass,Yang_2018,Klemenz_2020,Klemenz_2020b}. In general, a peculiar property of Dirac cones (DCs) is that they can merge into a so-called semi-Dirac (SD) point~\cite{Hasegawa_2006,Dietl_2008,Banerjee_2009,Montambaux_2009,Goerbig_2017}. These points are distinguished by an energy dispersion linear in one direction and parabolic in the other. They are particularly interesting for their topological~\cite{montambaux2018winding} and anisotropic transport properties~\cite{Real_2020}.

In this work, we investigated the spectral properties of a non-symmorphic wallpaper group lattice: the herringbone lattice (HL)~\cite{Conway_2008}, characterized by a pair of DCs. By breaking some of the system's symmetries, we can tune the position of these DCs within the FBZ to merge them into a SD one and eventually gap them.
We achieve similar results by modifying the internal strength of the hopping terms. Moreover, we show that this type of modification leads to the appearance of features similar to a system characterized by a set of parallel Su-Schrieffer-Heeger (SSH) chains~\cite{Heeger_1988,Jeon_2022,Li_2022,Herrera_2022}.
In Refs.~\cite{Jeon_2022,Li_2022}, the authors consider inclined SSH chains in a square lattice model. For a certain choice of parameters, the system presents a non-symmorphic symmetry~\cite{Jeon_2022,Li_2022}. However, the HL already possesses non-symmorphic symmetries for homogeneous hopping terms, which introduces additional constraints to the bands of the system. Specifically, it possesses band degeneracies along high-symmetry paths. We also find that the SD unfolds along a high-symmetry path for a specific set of parameters. This unfolding results in a nodal line presenting linear and  parabolic dispersion along parallel lines in the FBZ.

\section*{Results}
\subsection*{Model Hamiltonian and spectrum}\label{secOne}

\begin{figure}[!t]
	\includegraphics[width=\columnwidth]{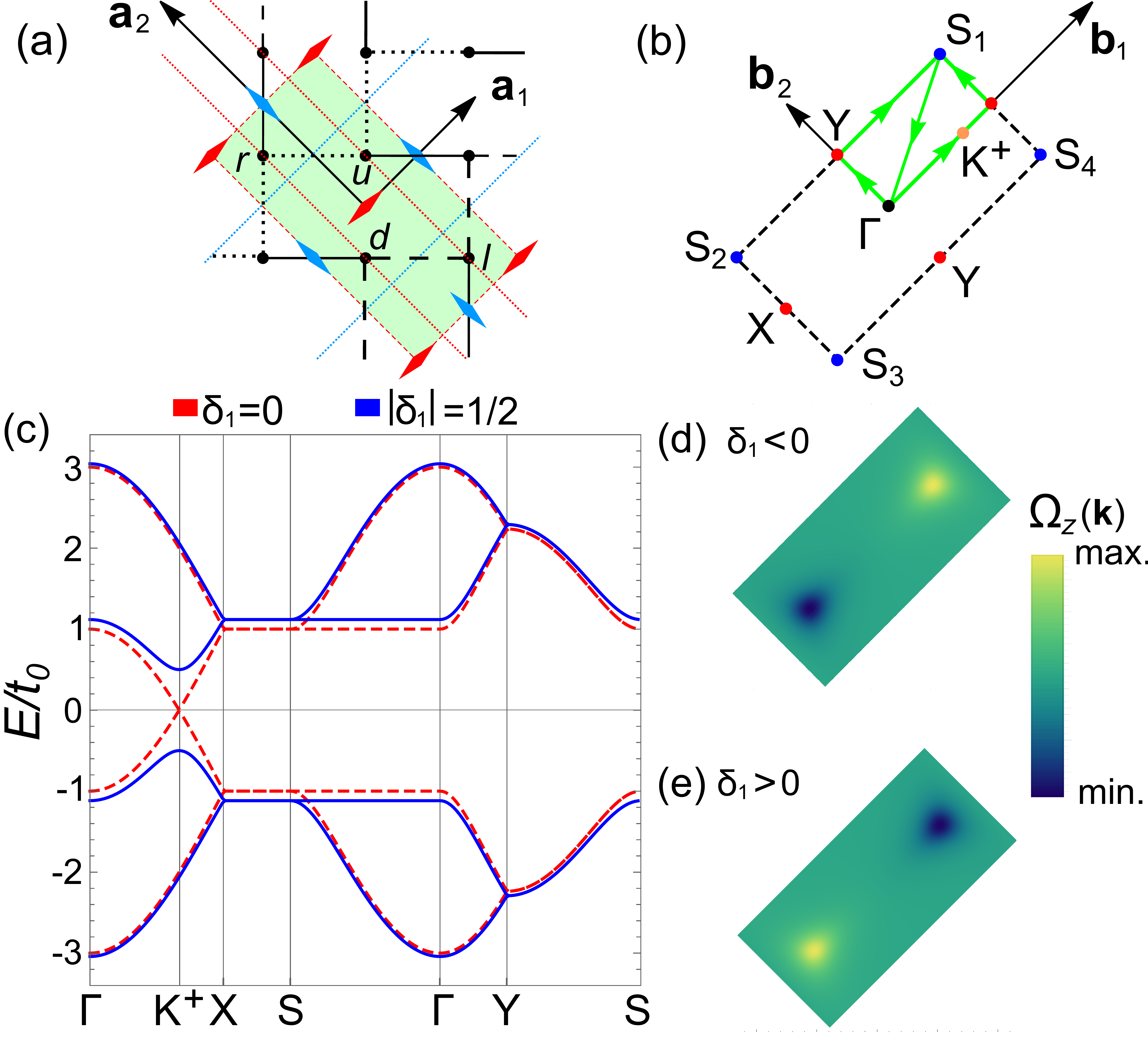}
	\caption{
 \textbf{Herringbone lattice in real, reciprocal and energy space: }(a) Herringbone lattice unit cell with the naming of the lattice sites $(r,d,u,l)$, direct lattice vectors $\mathbf{a}_1$ and $\mathbf{a}_2$, and the two sets of glides: the dashed blue lines correspond to $\mathcal{G}_{1\alpha}$ while the red ones to $\mathcal{G}_{2\alpha}$. The red and blue diamonds represent the $2a$ and $2b$ maximal Wyckoff positions, respectively. The Su-Schieffer-Heeger-like chains have been highlighted by two different densities of the dashing. (b) First Brillouin zone, reciprocal lattice vectors $\mathbf{b}_1$ and $\mathbf{b}_2$, and high-symmetry points, plus the position of the Dirac cones appearing in the irreducible Brillouin zone. (c) Energy spectrum ($E$ in units of $t_0$) along the high-symmetry path, for zero mass term (dashed red) and with $\mathcal{M}_1(\delta_1$) (solid blue). (e,f) Berry curvature inside the first Brillouin zone for $\delta_1=\pm 0.5$.\label{fig_1}} 
\end{figure}

The HL contains four sites in its unit cell, all with coordination number 3 | Fig.~\ref{fig_1}(a). The lattice vectors are $\mathbf{a}_1=(1,1)\,a_0$ and $\mathbf{a}_2=(-2,2)\,a_0$, where $a_0$ is the interatomic distance. We label each unit cell by $(m,n)=m\mathbf{a}_1+n\mathbf{a}_2$. The HL can be regarded as a square lattice with each site missing a link; thus, we classify the four sites in the unit cell  according to the remaining link: $r_{mn}$ and $l_{mn}$ along the horizontal direction with a link to the right and left, respectively, and $u_{mn}$ and $d_{mn}$ along the vertical direction with a link upwards and downwards, respectively.
Placing $s$-like orbitals, the HL nearest-neighbour tight-binding Hamiltonian reads
%
%
%%%%%%%%%%%
\begin{equation}
    \hat{\mathcal{H}}=\!\!\!\!\!\!\!\! \sum_{\substack{\langle m,n\rangle \\\langle\alpha,\alpha'\rangle\in
    \left\{r,d,u,l\right\} }}\!\!\!\!\left( t_{\alpha,\alpha'}\, c^\dagger_{\alpha,mn}c_{\alpha',mn}+\varepsilon_\alpha     c^\dagger_{\alpha,mm}c_{\alpha,mm} \right)\label{eq:ham}
\end{equation}
%%%%%%%%%
%
%
where $t_{\alpha,\alpha'}$ is the hopping amplitude between nearest-neighboring sites, and $c_{\alpha,mn}$($c^\dagger_{\alpha,mn}$) annihilates (creates) an electron on a lattice site $\alpha,\alpha'\in\{r,d,u,l\}$ at unit cell $(m,n)$. In Eq.~\eqref{eq:ham}, the symbol $\langle\ldots\rangle$ indicates nearest-neighbour lattice sites. The last term is the on-site $\varepsilon_\alpha$: we will use it to break some of the HL symmetries selectively. The hopping terms can be classified into intra-cell, where we only find horizontal terms, and inter-cell hoppings, where we find both horizontal and vertical terms. 

Neglecting the onsite energies and introducing $k_i=\mathbf{k}\cdot\mathbf{a}_i$, we can Fourier transform the HL Hamiltonian~\eqref{eq:ham} as follows $\hat{\mathcal{H}} = \sum_{\mathbf{k}} \hat{\Psi}^{\dagger}_{\mathbf{k}}h({\mathbf{k}})\hat{\Psi}_{\mathbf{k}}$, where
%
%
%%%%%%%%%%%%%
\begin{subequations}
\begin{align}
h({\mathbf{k}})&=
\begin{pmatrix} 
0 & q(\mathbf{k})\\
q(\mathbf{k})^\dagger & 0
\end{pmatrix},\label{eq_ham_matrix}\\
q(\mathbf{k})&=t_0\begin{pmatrix}\label{eq_q_fun}
1+\ee^{-\ii k_1} & \ee^{\ii k_2}\\
\ee^{-\ii k_1} & 1+\ee^{-\ii k_1} 
\end{pmatrix}.
\end{align}
\end{subequations}
%%%%%%%%%%%%%
%
%
We have set $t_{\alpha,\alpha'}=t_0$ for simplicity. With the following choice of basis $\Psi_\mathbf{k}=(\psi_r,\psi_d,\psi_u,\psi_l)^\text{T}$ the Hamiltonian in \eqref{eq_ham_matrix} fulfills the chiral-symmetry operator $\mathcal{C}=\tau_z\otimes \sigma_0$. This operator hints at an interpretation of the HL as two coupled SSH chains, each with its chiral symmetry~\cite{Jeon_2022,Li_2022}. The chains are formed by pairs of $r,u$ and $d,l$, respectively. The chains are connected along the horizontal direction via $u,d$ atoms and along the vertical direction via $r,l$ atoms | Fig.~\ref{fig_1}(a). Thus, in the operator $\mathcal{C}$, the $\tau_{x,yz}$ represent the intrachain degrees of freedom, while $\sigma_{x,y,z}$ represent the interchain ones. This will be addressed further in the text when breaking the various symmetries.
%
%
%%%%%%%%%%%%%%%
%%%%%%%%%%%%%%%
% 
%
The energy spectrum associated to Eq.~\eqref{eq_ham_matrix} reads:
%
% 
%%%%%%%%%%%%%
\begin{equation}
\mathcal{E}_{\alpha,\beta}(\mathbf{k})=\alpha t_0\sqrt{3\!+\!2\cos k_1\!+\!
4\beta\cos\left(\frac{k_1}{2}\right)\cos\left(\frac{k_2}{2}\right)}\label{eq_bandstruc}
\end{equation}
%%%%%%%%%%%%%
%
%
with $\alpha,\beta=\pm$. It presents four energy-symmetric bands, with several remarkable features | Fig.~\ref{fig_1}(c). To start, it displays DCs between bands 2 and 3 located at $\mathbf{K^\pm}=\pm\mathbf{b}_1/3$, along a high-symmetry path~\cite{Kruthoff_2017}. These cones are characterized by a $\pm\pi$ Berry phase. Additionally, we observe flat nodal lines along SXS lines and dispersive ones along SYS, both between bands 1\&2 and 3\&4. All these features are rooted in the symmetries of the HL: it belongs to the $pgg$ wallpaper group~\cite{Conway_2008}. This group is non-symmorphic, meaning that some symmetry operators do not leave any point of the space invariant since they include fractional translations along lattice vectors, called \emph{glide} symmetries. There are two sets of glides acting on different sites; thus, we name them with two indices, $\mathcal{G}_{i\alpha}$, where $i$ corresponds to the index of the lattice translation involved, and $\alpha=\{\text{A,B}\}$ depending on whether the $r$ lattice site is closest to the mirror | see Fig.~\ref{fig_1}(a).
Using Seitz, symbols~\cite{glazer2014seitz}, these glides are $\mathcal{G}_{1\alpha}=\left\{m_{01}|\tfrac{1}{2}0\right\}$ and $\mathcal{G}_{2\alpha}=\left\{m_{10}|0\tfrac{1}{2}\right\}$ | Fig.~\ref{fig_1}(a). The unit cell of the HL contains four maximal Wyckoff positions, and none of the glide planes go through them. However, it is remarkable how glides $\left\{\mathcal{G}_{2\alpha}\right\}$ go through the lattice sites, while $\left\{\mathcal{G}_{1\alpha}\right\}$ do not | Fig.~\ref{fig_1}(a). This affects how lattice sites transform under these symmetries: when applying the set of glides $\left\{\mathcal{G}_{1\alpha}\right\}$, no matter which symmetry operation is performed first (mirror or half translation), the lattice site falls on empty space, whereas for $\left\{\mathcal{G}_{2\alpha}\right\}$ the half translation already maps $r$ into $d$, and $u$ into $l$. These properties affect how the spectrum behaves after breaking $\left\{\mathcal{G}_{2\alpha}\right\}$ $vs$. $\left\{\mathcal{G}_{1\alpha}\right\}$. 
%
%
%%%%%%%%%%%%%%%%
\begin{figure}[!t]
	\includegraphics[width=\linewidth]{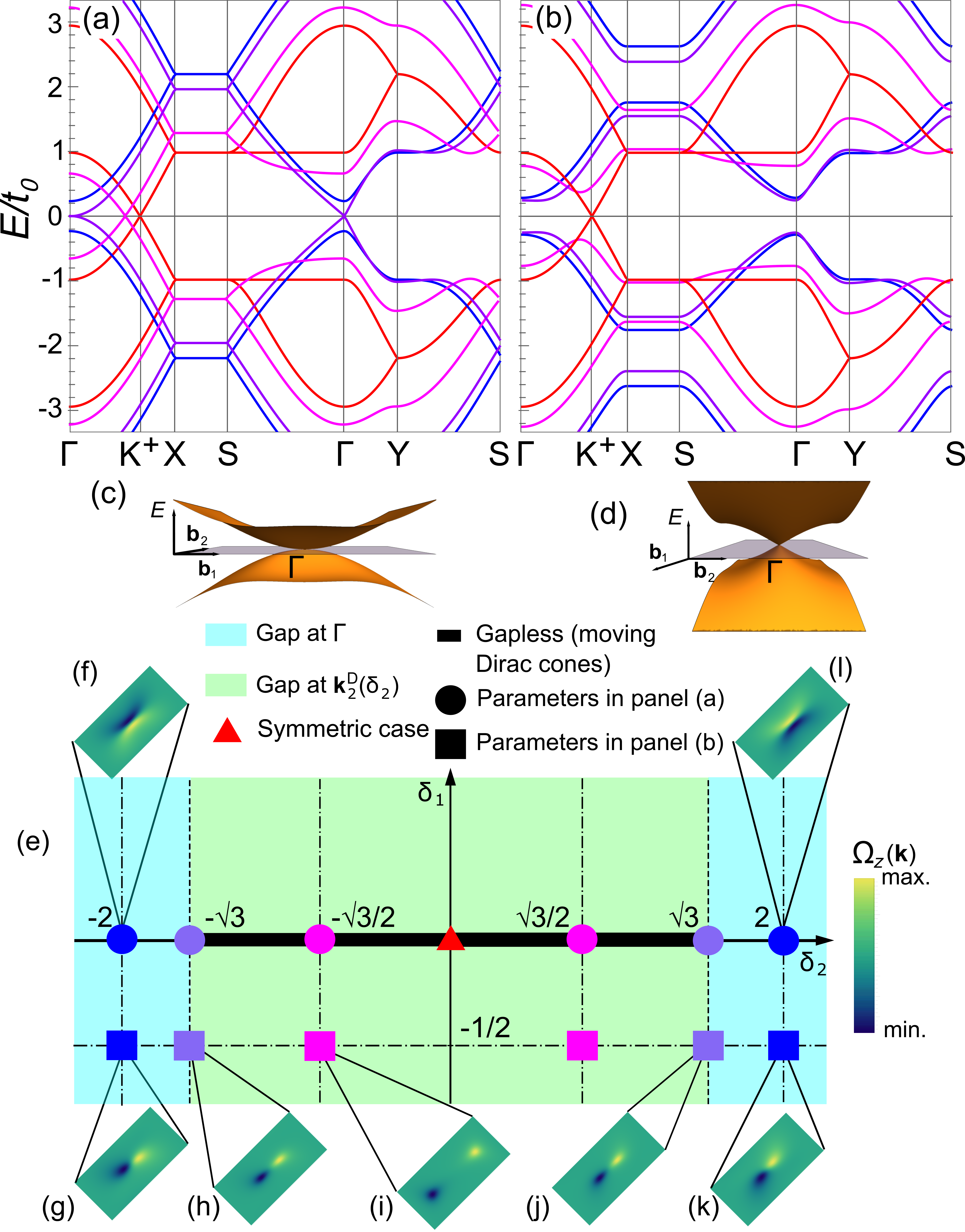}
	\caption{\textbf{Spectral properties of the Herringbone lattice under $\mathcal{M}_1(\delta_1)+\mathcal{M}_2(\delta_2)$: }(a) Energy spectrum ($E$ in units of $t_0$) with $\mathcal{M}_2(\delta_2)$ for several values of $\delta_2$. (b) Energy spectrum ($E$ in units of $t_0$) with $\mathcal{M}_\text{T}(\delta_1=-1/2,\delta_2)=\mathcal{M}_1(-1/2)+\mathcal{M}_2(\delta_2)$ for the same values of $\delta_2$ in (a). (c,d) Dispersion of bands 2\&3 displaying a semi-Dirac cone along two different orientations. (e) Phase diagram of the Herringbone lattice under $\mathcal{M}_\text{T}(\delta_1,\delta_2)$ according to the position of the gap. (f)-(j) Berry curvature distributions for different sets of parameters 
	($\delta_1,\delta_2$). The colorbar helps to distinguish peaks and valleys in the Berry curvature distributions}. \label{fig_2}
\end{figure}
%%%%%%%%%%%%%%%%
%
%
\subsection*{Strategies for tuning the Dirac cones}\label{secTwo}

Here, we show how to gap the cones, move them within the FBZ and eventually merge them into a SDC cone. All the onsite perturbations respect Tr $h(\mathbf{k})=0$. However, they all commute with $\mathcal{C}$, and as a consequence, this is not anymore a well-defined chiral symmetry. We will perturb the bands according to the SSH-like interpretation, meaning we will differentiate between chains and between lattice sites inside each chain in several ways.

To gap the cones, we fix opposite onsite energies at lattice sites $r$ and $u$, $d$ and $l$, \emph{i.e.}, $(\varepsilon_r,\varepsilon_d,\varepsilon_u,\varepsilon_l)=\delta_1(1,1,-1,-1)t_0$. This configuration differentiates between the inside each chain respecting the chiral symmetry, and it can be expressed by the $\left\{\mathcal{G}_{1\alpha}\right\}$-breaking mass term
%
%
%%%%%%%%%%%%
\begin{equation}
    \mathcal{M}_1(\delta_1)= \delta_1 \left(\tau_z\otimes\mathbb{I}_2\right)t_0.\label{eq_M1}
\end{equation}
%%%%%%%%%%%%
%
%
For $\delta_1=0$ (critical point), the spectrum is gapless and energy symmetric | see Fig.~\ref{fig_1}(c). Away from this value, the band structure splits into two gapped composite sets of bands. The rest of the spectral features (flat and degenerate lines) are shifted in energy, preserving the degeneracy.

The spectrum is symmetric with respect to $\delta_1=0$, but  the eigenfunctions behave differently after a change in sign of $\delta_1$: the Berry curvature~\cite{Xiao_2010,gradhand2012first,supplemental} reveals exchange of the charge at the K$^\pm$ points | Figs.~\ref{fig_1}(f) to~\ref{fig_1}(l). Having only two DCs inside the FBZ, the Berry curvature displays a dipolar distribution with a fixed length. Given $\{\mathcal{G}_{1_\alpha}\}$ involves the mirror $m_{01}$ in real space, the same mirror in reciprocal space is conserved, and the dipolar distribution is oriented along $\mathbf{b}_1$.

The breaking of $\left\{\mathcal{G}_{1\alpha}\right\}$ can also be studied from symmetry eigenvalues of $\left\{\mathcal{G}_{2\alpha}\right\}$ before and after the closing of the gap. We observe that the eigenvalues of bands 2 and 3 invert before and after the gap's closing, which reflects a band inversion | see Supplementary Note 2~\cite{supplemental}.

%
%
%%%%%%%%%%%%%%%%%%
\begin{figure}[!t]
	\includegraphics[width=\linewidth]{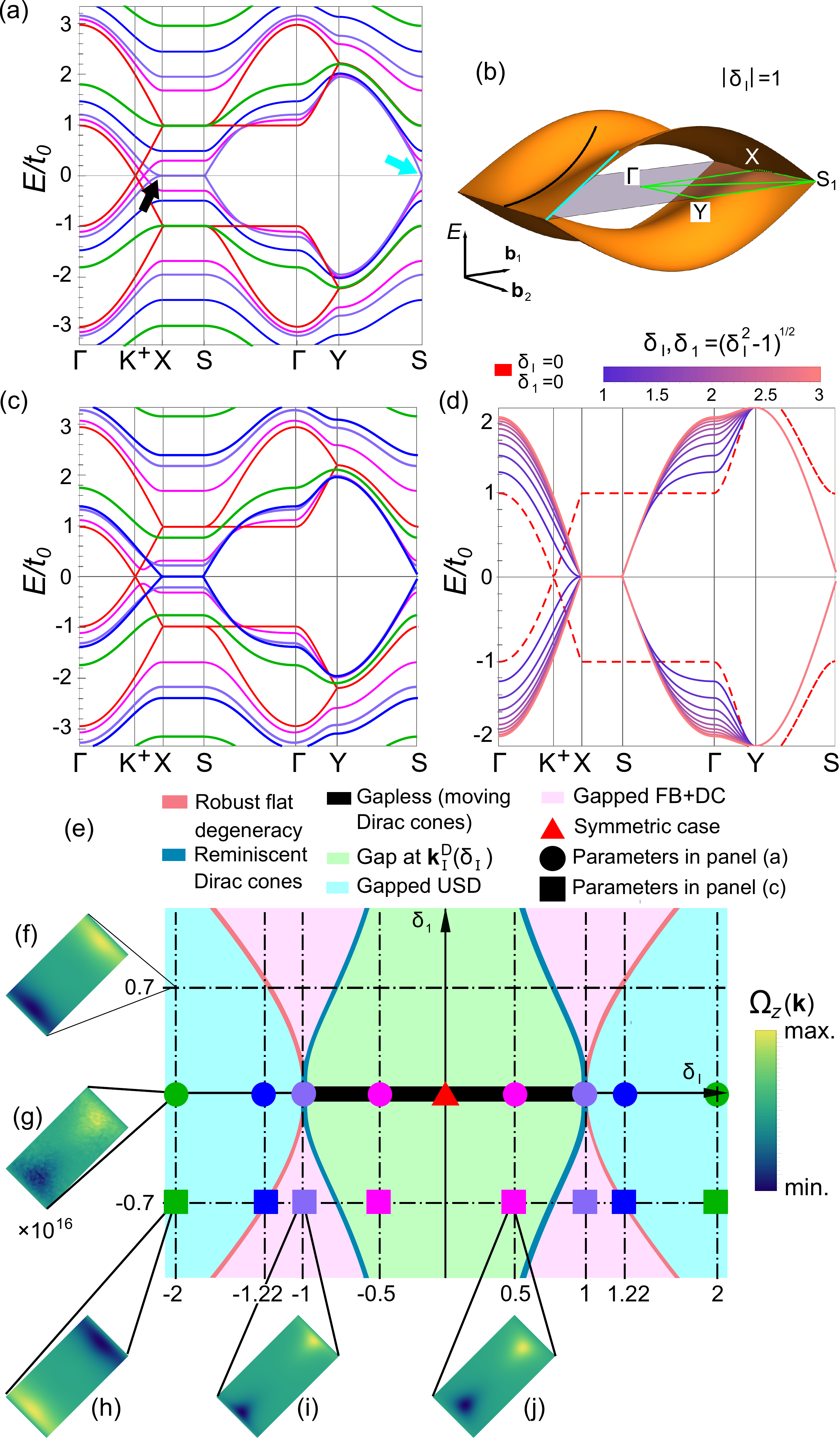}
	\caption{\textbf{Spectral properties of the Herringbone lattice under $\mathcal{M}_1(\delta_1)+\mathcal{M}_\text{I}(\delta_\text{I})$: }(a) Energy spectrum ($E$ in units of $t_0$) with $\mathcal{M}_\text{I}(\delta_\text{I})$ for different values of $\delta_\text{I}$. (b) Bands 2\&3 with the nodal line, where the parabolic/linear behavior around X/S points is highlighted. (c) Energy spectrum of the Herringbone lattice with $\mathcal{M}_\text{T}^\text{A}(\delta_\text{I},\delta_1=-0.7)=\mathcal{M}_\text{I}(\delta_\text{I})+\mathcal{M}_1(-0.7)$. (d) Robust nodal line. (e) Phase diagram of the Herringbone lattice with $\mathcal{M}_\text{T}^\text{A}(\delta_\text{I},\delta_1)$ according to the position of the gap. (f)-(j) Berry curvatures for the selected set of parameters. Band inversions occur for the positive value of $\delta_1$. The colorbar helps to distinguish peaks and valleys in the Berry curvature distributions.}\label{fig_3}
\end{figure}
\begin{figure}[!t]
	\includegraphics[width=\linewidth]{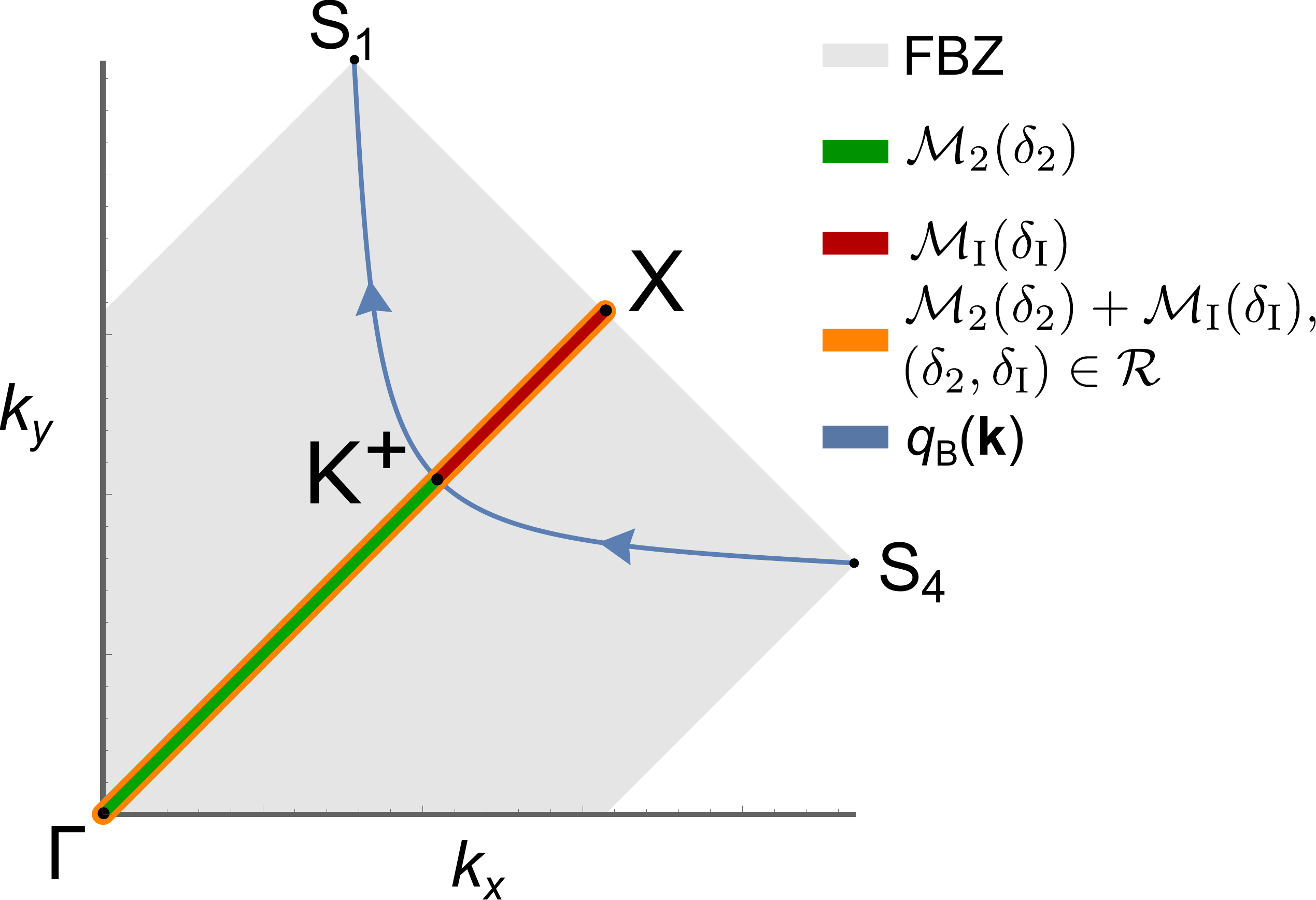}
	\caption{\textbf{Summary of the positions of the Dirac cones according to the different perturbations: } Representation of the all trajectories of the Dirac cones as they move across the first Brillouin zone as a function of the different perturbations (only showing positive values of $k_x,k_y$).}\label{fig_hyp} 
\end{figure}

%%%%%%%%%%%%%%%%%%
%
%
Next, we show the first strategy to move the DCs. We fix opposite onsite energies at lattice sites $r$ and $d$, $u$ and $l$, \emph{i.e.}, 
$(\varepsilon_r,\varepsilon_d,\varepsilon_u,\varepsilon_l)=\delta_2(1,-1,1,-1)t_0$. 
Now, we are differentiating between chains, by placing the same energy in chiral-symmetric lattice sites. This corresponds to the following  $\left\{\mathcal{G}_{2\alpha}\right\}$-breaking mass term 
%
%
%%%%%%%%%%%%%
\begin{equation}
\mathcal{M}_2(\delta_2)=\delta_2\left(\mathbb{I}_2\otimes\sigma_z\right)t_0.\label{eq_M2}
\end{equation}
%%%%%%%%%%%%%
%
%
We recover the unperturbed gapless phase for $\delta_2=0$. However, the band structure remains gapless within the interval $|\delta_2|\leq\sqrt{3}$ (degeneracy interval) but differs from the fully symmetric case. The band structure is symmetric with respect to a change in the sign of $\delta_2$, so as soon as $\delta_2\neq 0$ the DCs move away from K$^\pm$ towards $\Gamma$. We can find their position as a function of $\delta_2$ by solving $E_{3}[\delta_2,\mathbf{k}^\text{D}_2(\delta_2)]=0$, where $E_{3}$ is the third band, corresponding to $(\alpha,\beta)=(+,-)$. It yields:
%
%
%%%%%%%%%%%%
\begin{equation}
   \mathbf{k}^\text{D}_2(\delta_2)=\frac{1}{a_0}\arccos \left(\frac{1}{2}\sqrt{1+\delta_2}^2\right)\frac{\mathbf{b}_1}{|\mathbf{{b}_{1}|}}\label{eq_G2_kdelta}.
\end{equation}
%%%%%%%%%%%%
%
%
The motion of the DCs is shown in Fig.~\ref{fig_2}(a). For $|\delta_2|=\sqrt{3}$ (limits of the degeneracy interval), the DCs have shifted away from $\mathbf{K}^\pm$ (at $\delta_2=0$) merging at $\Gamma$ into a SDC. Figures~\ref{fig_2}(c) and~\ref{fig_2}(d) show the SDC at $\Gamma$ with the parabolic/linear behaviour explicitly displayed. For $|\delta_2|>\sqrt{3}$, the SD is gapped away, and the band structure again splits into two detached composite sets of bands, with the flat degeneracy untouched but the dispersive lifted up. The Berry curvature of this situation is shown in Figs.~\ref{fig_2}(f) and~\ref{fig_2}(l), where we observe a band inversion at $\Gamma$ depending on the sign of $\delta_2$. Figure~\ref{fig_hyp} displays the trajectory of the DCs for increasing $\delta_2$, starting at K$^\pm$ as DCs for $\delta_2=0$, shifting towards $\Gamma$ at $|\delta_2|=\sqrt{3}$ merging into a SDC.

There is another way in which we can gap the band structure, and it is by adding a $\mathcal{M}_1(\delta_1)$ term to the already existing $\mathcal{M}_2(\delta_2)$. The behavior of the bands under $\mathcal{M}_\text{T}(\delta_1,\delta_2)=\mathcal{M}_1(\delta_1)+\mathcal{M}_2(\delta_2)$ is shown in Fig.~\ref{fig_2}(b), where we have added a $\mathcal{M}_1(\delta_1=0.5)$ to the bands in Fig.~\ref{fig_2}(a). The unperturbed case has also been added as a guide to the eye. The overall effect is to gap the four bands everywhere, but more interestingly, to gap the DCs appearing at generic positions. With this total mass-term, we are able to shrink the Berry curvature dipolar distribution | Fig.~\ref{fig_2}(i) | as well as to visualize the Berry curvature of a gapped SDC, Figs.~\ref{fig_2}(h) and~\ref{fig_2}(j). Here we present some remarks after combining $\mathcal{M}_1(\delta_1)+\mathcal{M}_2(\delta_2)$: (i) a change in the sign of $\delta_1$ always produces a band inversion | Figs.~\ref{fig_2}(f) to~\ref{fig_2}(l) only display the Berry curvature for negative $\delta_1$, the ones for positive $\delta_1$ differ in an overall sign; (ii) the gap cannot be closed by using $\mathcal{M}_1(\delta_1)$ when $\delta_2$ falls outside the interval of degeneracy, so the gap at zero between the solid blue lines in Fig.~\ref{fig_2}(b) cannot be closed; (iii) the flat degeneracies along SXS remain flat but are completely lifted up by a nonzero value of $\delta_1$, so these two degeneracies are protected by both glides. 

Figure~\ref{fig_2}(e) displays all the gapped/metallic phases of the HL under $\mathcal{M}_\text{T}(\delta_1,\delta_2)$ depending on the position of the gap. The values of the parameters for bands in Figs.~\ref{fig_2}(a) and~\ref{fig_2}(b) are explicitly shown. At $\delta_1=0$ we find the phase diagram of $\mathcal{M}_2(\delta_2)$, which is an interval, and for $\delta_2=0$ we find the phase diagram of $\mathcal{M}_1(\delta_1)$, which is just the critical point.

The breaking of these glides can also be studied from symmetry eigenvalues of $\left\{\mathcal{G}_{1\alpha}\right\}$ for different values of $\delta_2$. We observe that the eigenvalues of bands 2 and 3 remain different for all values of $\delta_2$, so the crossing is still protected by the first set of glides (as soon as we add $\mathcal{M}_1(\delta_1)$, we gap the band structure). See Supplementary Note 2~\cite{supplemental}.
%
%
%%%%%%%%%%%%%%%%%%
\begin{figure}[!t]
	\includegraphics[width=\linewidth]{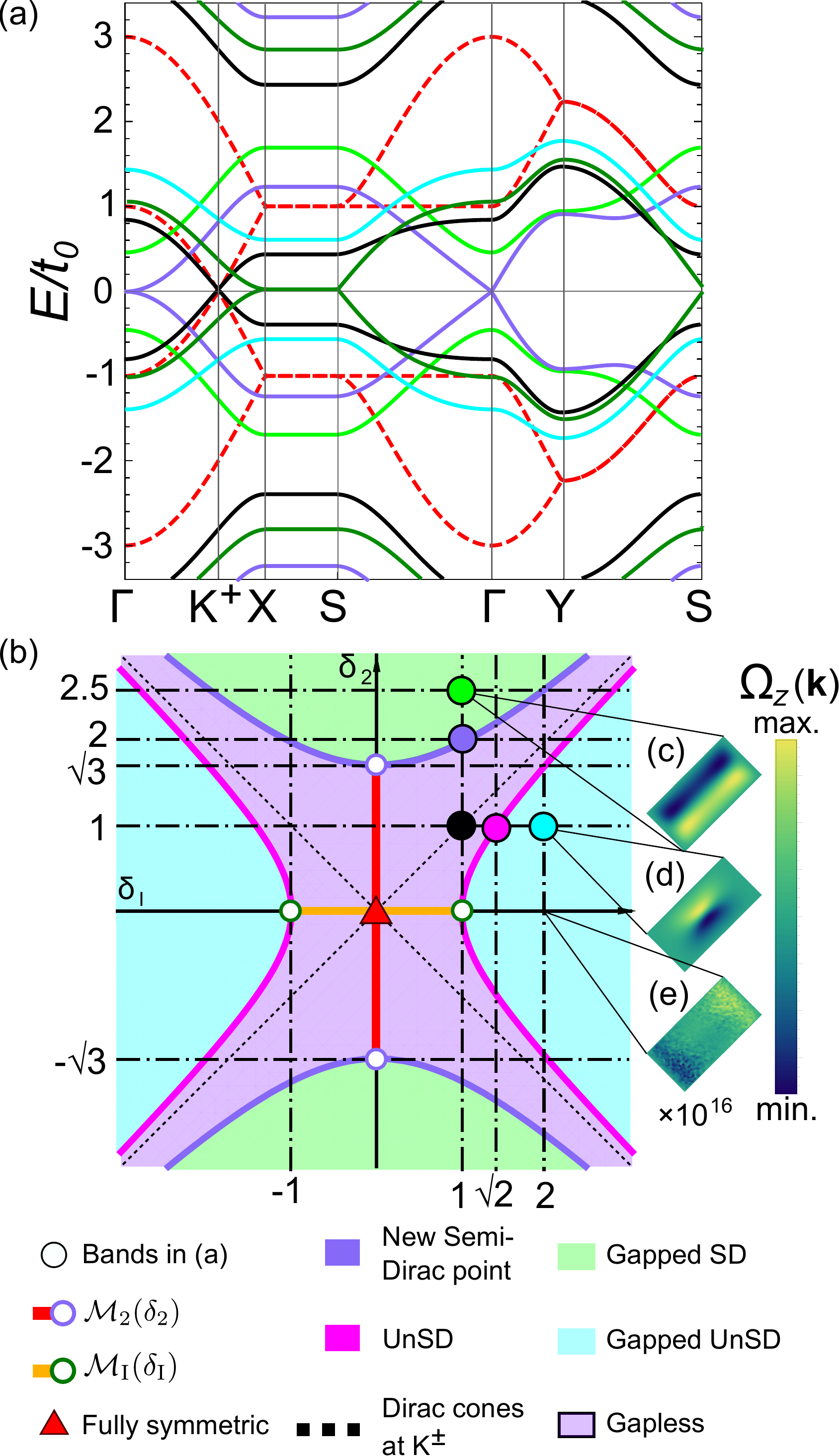}
	\caption{\textbf{Spectral properties of the Herringbone lattice under $\mathcal{M}_2(\delta_2)+\mathcal{M}_\text{I}(\delta_\text{I})$: }(a) Energy spectrum ($E$ in units of $t_0$) of the Herringbone lattice with $\mathcal{M}^\text{B}_\text{T}(\delta_2,\delta_\text{I})=\mathcal{M}_2(\delta_2)+\mathcal{M}_\text{I}(\delta_\text{I})$. The values of $(\delta_2,\delta_\text{I})$ are shown in (b). (b) Phase diagram of the Herringbone lattice with $\mathcal{M}^\text{B}_\text{T}(\delta_2,\delta_\text{I})$ according to the position of the gap. (c)-(e) Berry curvatures for the choices of parameters. The colorbar helps to distinguish peaks and valleys in the Berry curvature distributions}\label{fig_4}
\end{figure}
%%%%%%%%%%%%%%%%%%
%
%

Now, we present another strategy to tune the cones at different positions: simultaneously breaking both glides while respecting inversion symmetry. The origin of the unit cell represents the inversion center of the lattice, so in order to define this perturbation, we make $(\varepsilon_r,\varepsilon_d,\varepsilon_u,\varepsilon_l)=t_0\delta_\text{I}(1,-1,-1,1)$, or
%
%
%%%%%%%%%%%%%%
\begin{equation}
\mathcal{M}_\text{I}(\delta)= \delta_\text{I}\left(\tau_z\otimes\sigma_z\right)t_0\label{eq_MI}
\end{equation}
%%%%%%%%%%%%%%
%
%
This perturbation again commutes with the chiral operator $\mathcal{C}$. The spectrum displays common features between the two previous cases | Fig.~\ref{fig_3}(a). Bands 2\&3 touch for $|\delta_\text{I}|\leq 1$. Outside of it, the four bands are detached. For $|\delta_\text{I}|=\pm1$, bands 2 and 3 are degenerated at zero energy along SXS, forming a nodal line showing two different regimes: along the path X$\Gamma$X, the dispersion is locally parabolic around the X point. At the same time, it is locally linear around the S points along SYS. This is represented in Fig.~\ref{fig_3}(b) | the SDC is unfolded (UnSDC) along SXS, forming the nodal line. Shifting $\delta_\text{I}$ from -1 to 1, DCs appear from X (where the UnSDC is placed), shifting towards K$^\pm$ points at $\delta_\text{I}=0$ and going back to X. 
 
We  obtain  the analytical position of the DCs by solving $E_{3}[\delta_\text{I},\mathbf{k}^\text{D}_\text{I}(\delta_\text{I})]=0$. It yields:
%
%
%%%%%%%%%%%
\begin{equation}
    \mathbf{k}^\text{D}_\text{I}(\delta_\text{I})=\frac{1}{2 a_0}\arccos \left(-\frac{1+\delta_\text{I}^2}{2}\right) \frac{\mathbf{{b}_{1}}}{|\mathbf{{b}_{1}}|}\label{eq_kdeltaInv}
\end{equation}
%%%%%%%%%%%
%
%
Figure~\ref{fig_hyp} displays the trajectory of the DCs for increasing $\delta_\text{I}$. With this perturbation, we achieve the splitting of all degeneracies, which we have already seen after consecutively breaking both glide symmetries. However, here we are breaking both sets of glides at the same time while preserving inversion symmetry; thus, the results are different. Since we preserve time-reversal symmetry, the Berry curvature is zero in the FBZ for all $\delta_\text{I}$ values and bands.

We now combine the inversion-symmetric mass term $\mathcal{M}_\text{I}(\delta_\text{I})$ with the two previous mass terms, \emph{i.e.}, $\mathcal{M}_1(\delta_1)$ and $\mathcal{M}_2(\delta_2)$. We show in Fig.~\ref{fig_3}(c) the behavior of the band structure with total mass term $\mathcal{M}_\text{T}^\text{A}(\delta_\text{I},\delta_1)=\mathcal{M}_\text{I}(\delta_\text{I})+\mathcal{M}_\text{1}(\delta_1)$. This choice gaps the DCs at generic positions given by expression~\eqref{eq_kdeltaInv}, as shown in Fig.~\ref{fig_3}(c). Band inversions are detected using the distribution of the Berry curvature inside the FBZ, which has stretched with respect to Fig.~\ref{fig_1}(d) | Fig.~\ref{fig_3}(j). Robust degenerate flat bands along SXS are found by setting $\mathcal{M}_\text{T}^\text{A}(\delta_\text{I})=\mathcal{M}_\text{T}^\text{A}(\delta_\text{I},\delta_\text{I})=\mathcal{M}_\text{I}(\delta_\text{I})+\mathcal{M}_\text{1}(\pm\sqrt{\delta_\text{I}^2-1})$ for any $\delta_\text{I}>1$| see Fig.~\ref{fig_3}(d). 

Now that inversion has been broken, bands acquire a finite Berry curvature | Figs.~\ref{fig_3}(h) to~\ref{fig_3}(i). There is a line at which the flat features of the Berry curvature start to appear. In this situation, the gapped DCs merge into the band reaching the same value of the energy along the nodal line. We look for the relation between $(\delta_1, \delta_\text{I})$ that makes the energy at $\mathbf{k}_\text{I}^\text{D}(\delta_\text{I})$ equal to the energy along SXS since $\delta_1$ does not change the position of the DCs. We obtain:
%
%
%%%%%%%%%%
\begin{equation}
    \delta_1(\delta_\text{I})=\pm\frac{\sqrt{9 - 10 \delta_\text{I}^2 + \delta_\text{I}^4}}{4\delta_\text{I}}
\end{equation}
%%%%%%%%%%
%
%

All this information is displayed in the phase diagram of Fig.~\ref{fig_3}(e). The three coloured regions represent different gapped phases. As in the previous case, for $\delta_1=0$ we recover the phase diagram of $\mathcal{M}_\text{I}(\delta_\text{I})$, which is an interval. Setting $\delta_\text{I}=0$ we recover the phase diagram of $\mathcal{M}_1(\delta_1)$ (same as before). The Berry curvature distribution is shown for different sets of parameters. Along the line $\delta_\text{I}=-2$, the band inversion has been explicitly displayed since the Berry curvature changes sign without a gap closing. This is due to the fact that at $\delta_1=0, \delta_\text{I}>1$, inversion symmetry is recovered while the bands are fully gapped, and thus the Berry curvature is zero. Figure~\ref{fig_3}(g) has been rescaled to match the colorbar.

We study now the mass term $\mathcal{M}_\text{T}^\text{B}(\delta_\text{I},\delta_2)=\mathcal{M}_\text{I}(\delta_\text{I})+\mathcal{M}_\text{2}(\delta_2)$. Figure~\ref{fig_4}(a) shows the band structure for different values of $(\delta_2,\delta_\text{I})$. Figure~\ref{fig_4}(b) shows the phase diagram of this choice of mass term. It describes the physics of $\mathcal{M}_2(\delta_2)$ by making $\delta_\text{I}=0$, and the physics of $\mathcal{M}_\text{I}(\delta_\text{I})$ by making $\delta_2=0$. Starting from these two setups, we can expand the phase diagram in the following way. The vertical axis is delimited by the point where the gap closes at $\Gamma$ forming a SDC, so we solve $E_3[\delta_2,\delta_\text{I},\Gamma]=0$. We obtain:
%
%
%%%%%%%%%%
\begin{equation}
\delta_2(\delta_\text{I})=\pm\sqrt{3+\delta_\text{I}^2}
\end{equation}
%%%%%%%%%%
%
%

When $\delta_\text{I}\neq0$, the SDC is formed just by the two intermediate bands. This new SDC can be gapped by adding a $\mathcal{M}_1(\delta_1)$. However, with the three mass terms, the bands are no longer symmetric in energy by pairs since the four onsite energies are in general different, see Supplementary Note 3~\cite{supplemental}.

We can split the new SDC into DCs moving across the whole FBZ. The position of the DCs is now governed by:
%
%
%%%%%%%%%%%
\begin{equation}
    \mathbf{k}^\text{D}_\text{I+2}(\delta_\text{I},\delta_2)=\frac{1}{2 a_0}\arccos \left(\frac{\delta_2^2-\delta_\text{I}^2-1}{2}\right)\frac{\mathbf{{b}_1}}{|\mathbf{{b}}_1|}\label{eq_kdeltaInv2}.
\end{equation}
%%%%%%%%%%%
%
%
By making $\delta_2=\pm|\delta_\text{I}|$ the new DCs always locates at K$^\pm$. Figure~\ref{fig_hyp} displays the trajectory of these DCs moving across the whole FBZ, from X, where a SDC splits into two DCs that move towards $\Gamma$, merging again into a SDC.

In order to expand the horizontal axis, at $(\delta_2,\delta_\text{I})=(0,\pm1)$, the gap closes along SXS in an UnSDC, so we solve $E_3[\delta_2,\delta_\text{I},X]=0$ to find the extension of the UnSDC for non-zero $\delta_2$. We obtain:
%
%
%%%%%%%%%%
\begin{equation}
\delta_2(\delta_\text{I})=\pm\sqrt{\delta_\text{I}^2-1}
\end{equation}
%%%%%%%%%%
%
%
The band structure is gapless at zero energy inside the region
%
%
%%%%%%%%%%
\begin{equation}
    \mathcal{R}=\left\{ (\delta_2,\delta_\text{I}):\,\delta_2^2- \delta_\text{I}^2 <|3| \cup \delta_\text{I}^2 - \delta_2^2 < 1\right\}
\end{equation}
%%%%%%%%%%
%
%
With this, we complete the phase diagram and our study of the onsite energies. In Supplementary Note 1~\cite{supplemental}, we present a low-energy theory~\cite{koshino2013electronic} for all the cases studied so far.

%
%
%%%%%%%%%%%%%%%%%%
\begin{figure}[!t]
	\includegraphics[width=\linewidth]{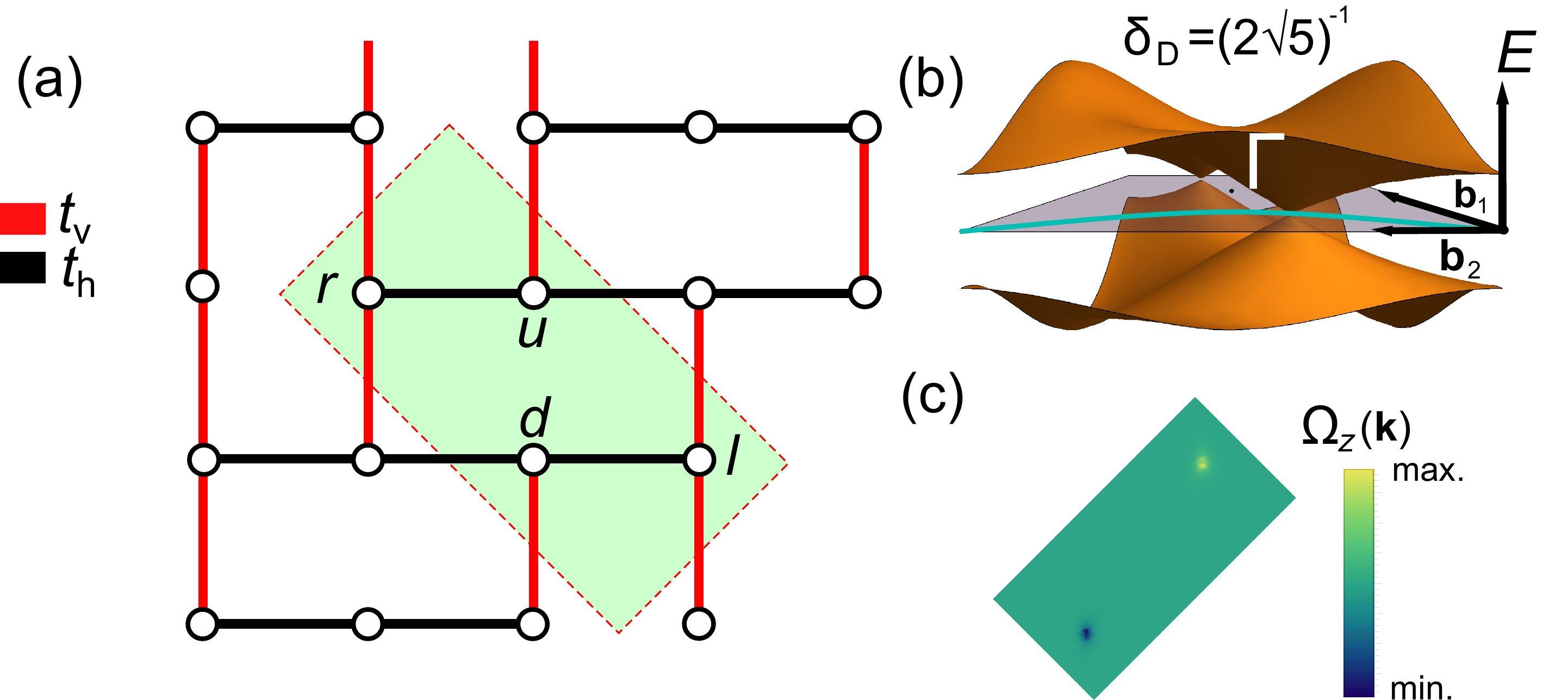}
	\caption{\textbf{Spectral properties of the dimerized Herringbone lattice: }(a) Sketch of the chosen dimerized implementation of the Herringbone lattice in real space. (b) Bands 2\&3 of the dimerized Herringbone lattice for $\delta_\text{D} = (2\sqrt{5})^{-1}$. The Dirac cones are now inside the first Brillouin zone and not along a high symmetry point or line. Their position as a function of $\delta_\text{D}$ is displayed. (c) Berry curvature for the dimerized Herringbone lattice after gaping the Dirac cones with $\mathcal{M}_1(\delta_1)$. The colorbar helps to distinguish peaks and valleys in the Berry curvature distributions}\label{fig_6}
\end{figure}
%%%%%%%%%%%%%%%%%%
%
%
Finally, we present a completely different strategy for tuning the DCs. It is based on differentiating between the horizontal and vertical hopping amplitudes in a \emph{breathing} form  $t_{h/v}=t_0(1\pm\delta_\text{D})$~\cite{Liu_2017,Li_2022}. There are several other choices to distort the system that also preserve some of the glides, but we studied the breathing one since it is the one yielding corner modes in certain geometries~\cite{Herrera_2022}. This breathing distortion breaks both glides while preserving inversion symmetry. We add to Eq.~\eqref{eq_q_fun} the matrix:
%
%
%%%%%%%%%%%
\begin{equation}
    q_\text{B}(\mathbf{k})=t_0\delta_\text{D}
\begin{pmatrix}
   1- \ee^{-\ii k_1} & - \ee^{\ii k_2} \\
  \ee^{-\ii k_1} & 1- \ee^{-\ii 
 k_1} \\
\end{pmatrix}
 \label{eq_kdeltaDim}.
\end{equation}
%%%%%%%%%%%
%
%
After diagonalizing the Hamiltonian, SDCs appear at the S$_{2,4}$ points for $\delta_\text{D}=-1/\sqrt{5}$. For increasing $\delta_\text{D}$, these cones split into DCs moving out from the S points towards the \textbf{K}$^\pm$ for $\delta_\text{D}=0$, where the fully symmetric case is recovered. For positive $\delta_\text{D}$, the cones keep moving continuously until they reach S$_{1,3}$, where they merge into SDC points with the ones coming from the neighbouring reciprocal unit cells. The trajectory of the cones is quasi-hyperbolic. If now we add onsite potentials according to~\eqref{eq_M1} and~\eqref{eq_M2}, the overall effect is to gap the cones at the arbitrary positions along the quasi-hyperbolic curve. This translates into an arbitrary orientation and length of the Berry curvature dipolar distribution as shown in Fig.~\ref{fig_6}(c). We show the trajectory of the DCs as a function of $\delta_\text{D}$ in Fig.~\ref{fig_hyp}.
%
%
%%%%%%%%%%%%%%%%%
\begin{table}
      \centering
    \begin{tabular}{cccc}
    \toprule
  + & $\mathcal{M}_1(\delta_1)$ & $\mathcal{M}_2(\delta_2)$ & $\mathcal{M}_\text{I}(\delta_\text{I})$ \\
   \midrule
  $\mathcal{M}_1(\delta_1)$ & Gap at K$^\pm$ &\begin{tabular}{c}$\sbullet\,$gapped SDC\\ $\sbullet\,$Shrunken\\ BC dist.\end{tabular}& \begin{tabular}{c}$\sbullet\,$gapped UnSDC\\ $\sbullet\,$Stretched BC dist.\\ $\sbullet\,$Robust flat\\ degeneracy\end{tabular} \\
  \midrule
  $\mathcal{M}_2(\delta_2)$ &  & 
    \begin{tabular}{@{}c@{}}$\sbullet\,$SDC \\ $\sbullet\,$Cones at \\$\mathbf{k}_2^\text{D}(\delta_2)$\end{tabular}
  & \begin{tabular}{@{}c@{}}$\sbullet\,$new SD +UnSD \\ $\sbullet\,$Cones inside \\$\mathcal{R}$ at $\mathbf{k}_{\text{I}+2}^\text{D}(\delta_\text{I})$\end{tabular}\\
  \midrule
  $\mathcal{M}_\text{I}(\delta_\text{I})$&&&\begin{tabular}{@{}c@{}}$\sbullet\,$UnSDC \\ $\sbullet\,$Cones at $\mathbf{k}_\text{I}^\text{D}(\delta_\text{I})$\end{tabular}\\
  \bottomrule
\end{tabular}
  \caption{Summary of the main features of the combination of mass terms. Only the upper triangle has been filled not to overload the table.\label{tab_summary}}
\end{table}

%
%
%%%%%%%%%%%%%%%%%
%%%%%%%%%%%%%
%
%
In this work, we have shown how to tune the position of the DCs of the non-symmorphic HL. We have proposed several strategies for merging the cones into a SDC one and eventually opening an energy gap in the system. We have summarized all the possible positions of the DCs within the BZ in Fig.~\ref{fig_hyp}. In addition to tuning the position of the DCs, we can manipulate the orientation of the Berry curvature dipolar distribution, from being parallel to the reciprocal lattice vectors to having a generic length and orientation. Table~\ref{tab_summary} summarizes the action of the onsite perturbations studied so far in the system. As mentioned above, the combination of breathing plus onsite only gaps the spectrum, so it has not been added to the table.

\section*{Conclusions}\label{secThree}
\begin{figure}[!t]
	\includegraphics[width=\columnwidth]{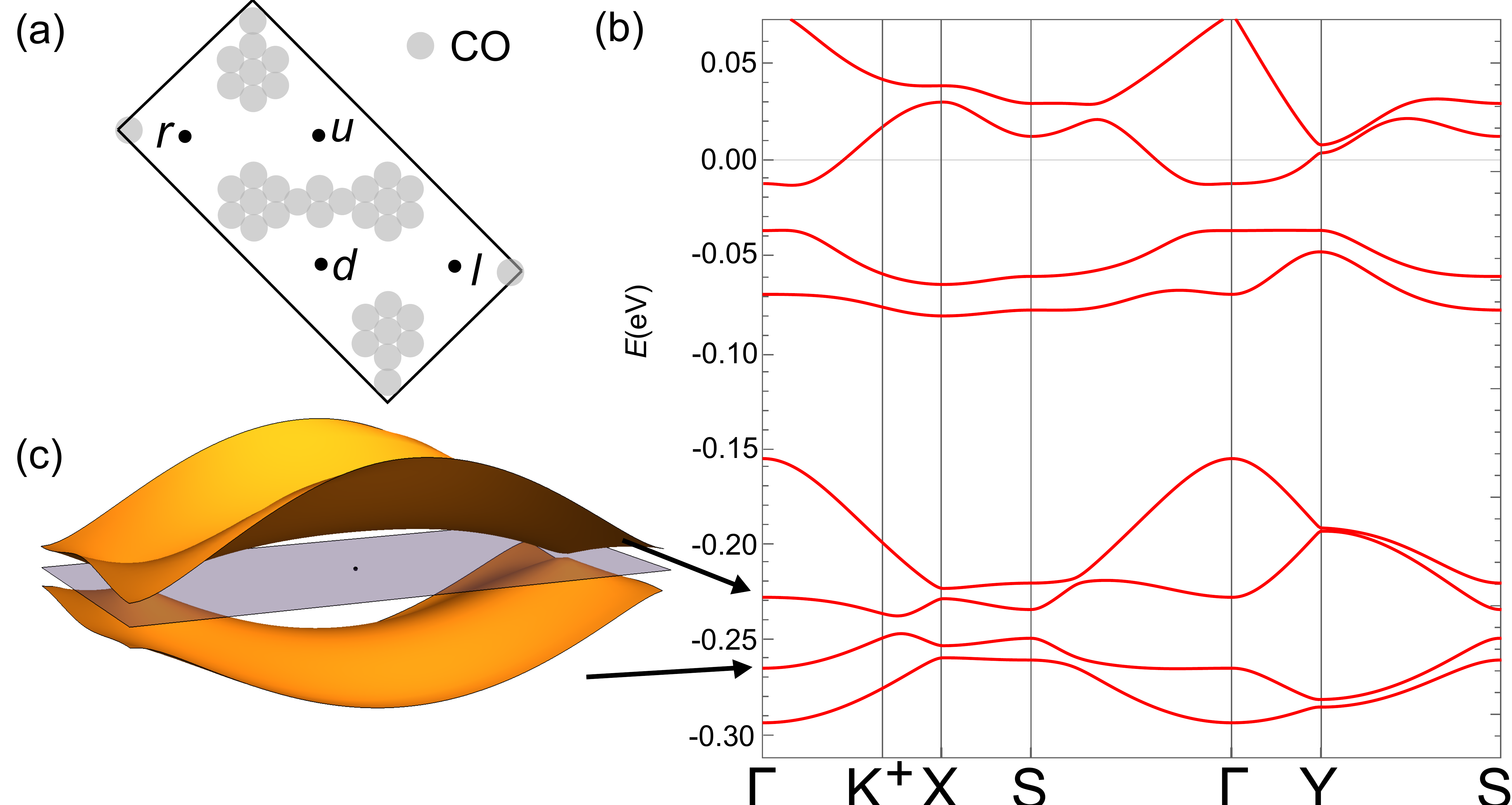}
	\caption{\textbf{Spectral properties of the simulated Herringbone lattice: }(a) Proposal for the unit cell of the Herringbone lattice in the nearly-free electron simulator. (b) Energy spectrum along a high-symmetry path for the Herringbone lattice according to the proposal. (c)~Bands 2\&3 of the proposal, underlining the presence of the Dirac cones outside a high-symmetry path.\label{fig_7}} 
\end{figure}

 The moving and the merging of the DCs has been already experimentally observed in black phosphorous~\cite{Fei_2015,Kim_2017}:  a 2D layered material characterized by non-symmorphic symmetries. We propose here a realization of the HL within the synthetic platform known as the artificial electron lattice~\cite{khajetoorians2019creating}. Here the two-dimensional electron gas hosted on the (111) surface state of Cu is confined to a potential well designed with a set of CO molecules, which are placed with atomic precision at certain positions with the help of the tip of a scanning tunnelling microscope~\cite{Gomes_2012, Slot_2017,Kempkes_2018,Freeney_2020,Kempkes_2019,gardenier2020p,stilp2021artificial_atoms,Herrera_2022}. We present the design of the HL in Fig.~\ref{fig_7}(a). Symmetry plays a crucial role: if the space groups of the substrate and the simulated lattice have common generators  (one is a subgroup of the other), then the electronic structure of the lattice is very well recovered. However, if this condition is not met, it is more difficult to describe a lattice with this technique~\cite{gardenier2020p}. In our case, we expect something similar for our proposal. First of all, Fig.~\ref{fig_7}(b) shows the lower bands obtained for the unit cell depicted in Fig.~\ref{fig_7}(a). Only the lowest four bands come from the inner electronic levels of the artificial electronic lattice, and so they represent the bands closer to our spectrum presented in the results section. Figure~\ref{fig_7}(c) shows bands 2\&3 inside the FBZ, and we can see how two DCs appear at opposite $k$ points. From the discussion in the previous section, we can already see that the proposed unit cell will show some dimerization plus some onsite energies that will return to the position of the DCs. To fit these bands to a tight-binding Hamiltonian, next-nearest neighbours may be included, and even longer range hoppings, since the nearly-free electron method does not involve atomic orbitals or species, nor chemical bonds between them. The lattice sites are built with artificial interacting quantum dots (also known as artificial atoms~\cite{stilp2021artificial_atoms}) connected by hopping amplitudes which are always long-range and modelled by potential wells or barriers. We present further details about the calculation of the spectrum in the Method section.
In conclusion, we have presented the spectral properties of a 2D non-symmorphic lattice. We have shown that the system is characterized by two DCs along a high-symmetry line that can either gap or move within the FBZ. We also achieve the possibility of merging these DCs into a SDC one or, in a special case, into an UnSDC that respects the nodal line degeneracy imposed by a glide symmetry. 

%
%
%%%%%%%%%%%%%%%
%%%%%%%%%%%%%%%
%
%

\section*{Methods}
\subsection*{Berry curvature}\label{Berry_curvature}
We have classified the topological character of the bands below the Fermi level using the Berry curvature as a topological marker. It is important to note that there are always two bands below zero energy that can be degenerated or not. To evaluate the Berry curvature we made use of the Kubo formula~\cite{gradhand2012first}, provided the occupied set of bands is well separated from the unoccupied one:
%
%
%%%%%%%%%%%%%%%
\begin{equation}
    \Omega_{\gamma}(\mathbf{k})=\ii\epsilon_{\alpha\beta\gamma}\sum_{n}\sum_{m}\frac{\langle u_n|v_\alpha|u_m\rangle\langle u_m|v_\beta|u_n\rangle}{(E_m-E_n)^2}\,,
\end{equation}
%%%%%%%%%%%%%%
%
%
where $v_\mu=\partial_\mu \mathcal{H}$ are electron velocity along the $\mu$-direction, $|u_n\rangle$ are the periodic parts of the Bloch wave functions, and $\epsilon_{\alpha\beta\gamma}$ is the Levi-Civita antisymmetric tensor.

\subsection*{Energy spectrum for the artificial electron lattice}

The solution to the bulk problem for the artificial electron lattice is obtained within the nearly-free electron method~\cite{khajetoorians2019creating,park2009making,Herrera_2022}. We model each CO molecule as a cylindrical potential barrier placed at position $r_0$:
%
%
%%%%%%%%%%%
\begin{align}\label{cylinder}
V(r-r_0) & = \begin{cases} V_0 & r\leq r_0 \\
  0 & r > r_0 \end{cases}.
\end{align}
%%%%%%%%%%%
%
%
The complete lattice $V_\text{latt}(\mathbf{r})$ is given by the superposition of such potential barrier in Eq.~\eqref{cylinder}. Then, we expand the periodic potential in Fourier components
%
%
%%%%%%%%%%%%
\begin{equation}\label{eq:fourierseries}
V_{\rm latt}(\mathbf{r})=\sum_{\mathbf{G}} V_{\mathbf{G}}\text{e}^{\text{i} \mathbf{G}\cdot\mathbf{r}},
\end{equation}
%%%%%%%%%%%%
%
%
which are 
%
%
%%%%%%%%%%%%
\begin{align}\label{eq:fouriercomp}
V_{\mathbf{G}} &= \frac{1}{\sqrt{2\pi}}\int_\text{unit cell} d\mathbf{r}\,\, \text{e}^{-\text{i} \mathbf{G}\cdot \mathbf{r}} V_{\rm latt}(\mathbf{r})\, \nonumber\\
&=V_0r_0\sum_{\alpha_0} \text{e}^{-\text{i} \mathbf{G}\cdot\mathbf{R}^{\alpha}}  \frac{\mathcal{J}_1(|\mathbf{G}|r_0)}{|\mathbf{G}|},
\end{align} 
%%%%%%%%%%%%%
%
%
where $\mathbf{R}^{\alpha}$ are the positions of the molecules inside the unit cell, and $\mathcal{J}_1(|\mathbf{G}|r_0)$ is the Bessel function of the first kind. \\
The stationary Schr\"odinger equation is transformed into a set of linear equations for the coefficients $\{c_{n,\mathbf{q}}\}$ | which expand the wave function in plane waves | and the energy of the system,
%
%
%%%%%%%%%%%%
\begin{equation}\label{eq_scheq}
{\left(\frac{\hbar^2 \left|\mathbf{q}-\mathbf{G}\right|^2}{2m}-\mathcal{E}\right) c_{\mathbf{q}-\mathbf{G}} + \sum_{\mathbf{G'}}  V_{\mathbf{G}'-\mathbf{G}} 	\, c_{\mathbf{q}-\mathbf{G}}=0},
\end{equation}
%%%%%%%%%%%%
%
%
solving this equation for the energy $\mathcal{E}$ results in the energy spectrum artificial electron lattice in the nearly free electron approximation. From the coefficients $\{c_{n,\mathbf{q}}\}$ that are used to obtain the periodic part of the wave function as
%
%
%%%%%%%%%%%%%%
\begin{equation*}
    u_{n,\mathbf{q}}(\mathbf{r})=\sum_\mathbf{G} \ee^{\ii \mathbf{G}\cdot\mathbf{r}} c_{n,\mathbf{q}-\mathbf{G}}.
\end{equation*}
%%%%%%%%%%%%%%
%
%

\section*{Code availability}
The codes that were employed in this study are available from the authors on reasonable request.

\def\bibsection{\section*{\refname}} 
\bibliography{non-bibliography.bib} 

\begin{acknowledgments}
We acknowledge valuable discussions with  Cristiane de Morais Smith, Alessandro De Martino,  Omjyoti Dutta, Aitzol Garcia-Extarri, Duy Hoang Minh Nguyen,  Giandomenico Palumbo, Vittorio Peano, and Ingmar Swart.  The work of M.A.J.H. and D.B. is supported by Ministerio de Ciencia e Innovaci\'on  (MICINN) through Project No.~PID2020-120614GB-I00 (ENACT) and by the Transnational Common Laboratory $Quantum-ChemPhys$ (D.B.). Additionally, D.B. acknowledges the funding from the Basque Government's IKUR initiative on Quantum technologies (Department of Education). 
\end{acknowledgments}

\section*{Competing interests}
The authors declare no competing interests.
\section*{Author contributions}
M. A. J. Herrera wrote the code for studying the lattice with the help and supervision of D. B.

  \cleardoublepage
\numberwithin{equation}{section}\setcounter{figure}{0}\global\long\def\thefigure{S\arabic{figure}}
\global\long\def\thesection{S\arabic{section}}
\global\long\def\thesubsection{\Alph{subsection}}
\setcounter{section}{0}

\begin{widetext}
\begin{center}
\textbf{\large{}Supplemental Material for ``Tunable Dirac points in a two-dimensional non-symmorphic wallpaper group lattice"}{\large{} }
\par\end{center}{\large \par}
\end{widetext}

\section{Supplementary Note 1: Low-energy description of the system Hamiltonian in the various phases}
In the following, we illustrate a general method that we have combined with the Taylor expansion to obtain the low-energy expression for the system Hamiltonian in the various phases presented in the main text.\\

\subsection{Decimation}
We consider  the energy eigenvalue equation, and consider separate
blocks in the $4\times4$ Hamiltonian corresponding to
low-energy $\lambda = \left( \psi_{\text{A}_1} , \psi_{\text{B}_2} \right)^\text{T}$
and dimer $\Delta = \left( \psi_{\text{A}_2} , \psi_{\text{B}_1} \right)^\text{T}$ components:
%
%
%%%%%%%%%
\begin{align}\label{tc}
\begin{pmatrix}
    h_{\lambda} & u \\
    u^{\dagger} & h_{\Delta} \\
  \end{pmatrix}
  \begin{pmatrix}
    \lambda \\
    \Delta \\
  \end{pmatrix}
= E 
  \begin{pmatrix}
    \lambda \\
    \Delta \\
  \end{pmatrix} \, , 
\end{align}
%%%%%%%%%
%
%
The second-row of~\eqref{tc} allows the dimer components to be expressed in terms of
the low-energy ones:
%
%
%%%%%%%%%%%%
\begin{eqnarray}
\Delta = \left( E - h_{\Delta} \right)^{-1} u^{\dagger} \lambda \, , \label{high}
\end{eqnarray}
%%%%%%%%%%%%
%
%
Substituting this into the first-row of~\eqref{tc} gives an effective eigenvalue
equation that is written solely for the low-energy components:
%
%
%%%%%%%%%
\begin{align*}
\left[ h_{\lambda} + u \left( E - h_{\Delta} \right)^{-1} u^{\dagger} \right] \lambda =& E \lambda \, , \\
\left[ h_{\lambda} - u h_{\Delta}^{-1} u^\dag \right] \lambda \approx& E \mathcal{S} \lambda \, ,
\end{align*}
%%%%%%%%%%%%
%
%
where $\mathcal{S} = 1 + u h_{\Delta}^{-2} u^\dag$. The second equation is accurate up to linear terms in $E$.
Finally, we perform a transformation $\Phi = \mathcal{S}^{1/2} \lambda$:
%
%
%%%%%%%%%
\begin{align} \label{heff}
\left[ h_{\lambda} - u h_{\Delta}^{-1} u^{\dagger} \right] {\cal S}^{-1/2} \Phi \approx& E {\cal S}^{1/2} \Phi \, , \nonumber \\
\mathcal{S}^{-1/2} \left[ h_{\lambda} - u h_{\Delta}^{-1} u^{\dagger} \right] \mathcal{S}^{-1/2} \Phi \approx& E \Phi \, .
\end{align}
%%%%%%%%%
%
%
This transformation ensures that the normalisation of $\Phi$ is consistent with that of the original states:
%
%
%%%%%%%%
\begin{align*}
\Phi^{\dagger}  \Phi = \lambda^{\dagger} \mathcal{S} \lambda
=& \lambda^{\dagger} \left( 1 + u h_{\Delta}^{-2} u^{\dagger} \right) \lambda \, , \\
\approx& \lambda^{\dagger} \lambda + \Delta^{\dagger} \Delta \, ,
\end{align*}
%%%%%%%
%
%
where we used Eq.~\eqref{high} for small $E$: $\chi \approx - h_{\chi}^{-1} u^{\dagger} \lambda$.
Thus, the effective Hamiltonian for low-energy components is given by Eq.~\eqref{heff}:
%
%
%%%%%%%%%
\begin{subequations}\label{trans}
\begin{align}
H^{(\text{eff})} \approx& \mathcal{S}^{-1/2}
\left[ h_{\lambda} - u h_{\chi}^{-1} u^{\dagger} \right] \mathcal{S}^{-1/2} \, , \label{trans1} \\
\mathcal{S} =& 1 + u h_{\chi}^{-2} u^{\dagger} \, . \label{trans2}
\end{align}
\end{subequations}

\subsection{Low-energy Hamiltonians}

\begin{figure*}
	\includegraphics[width=\linewidth]{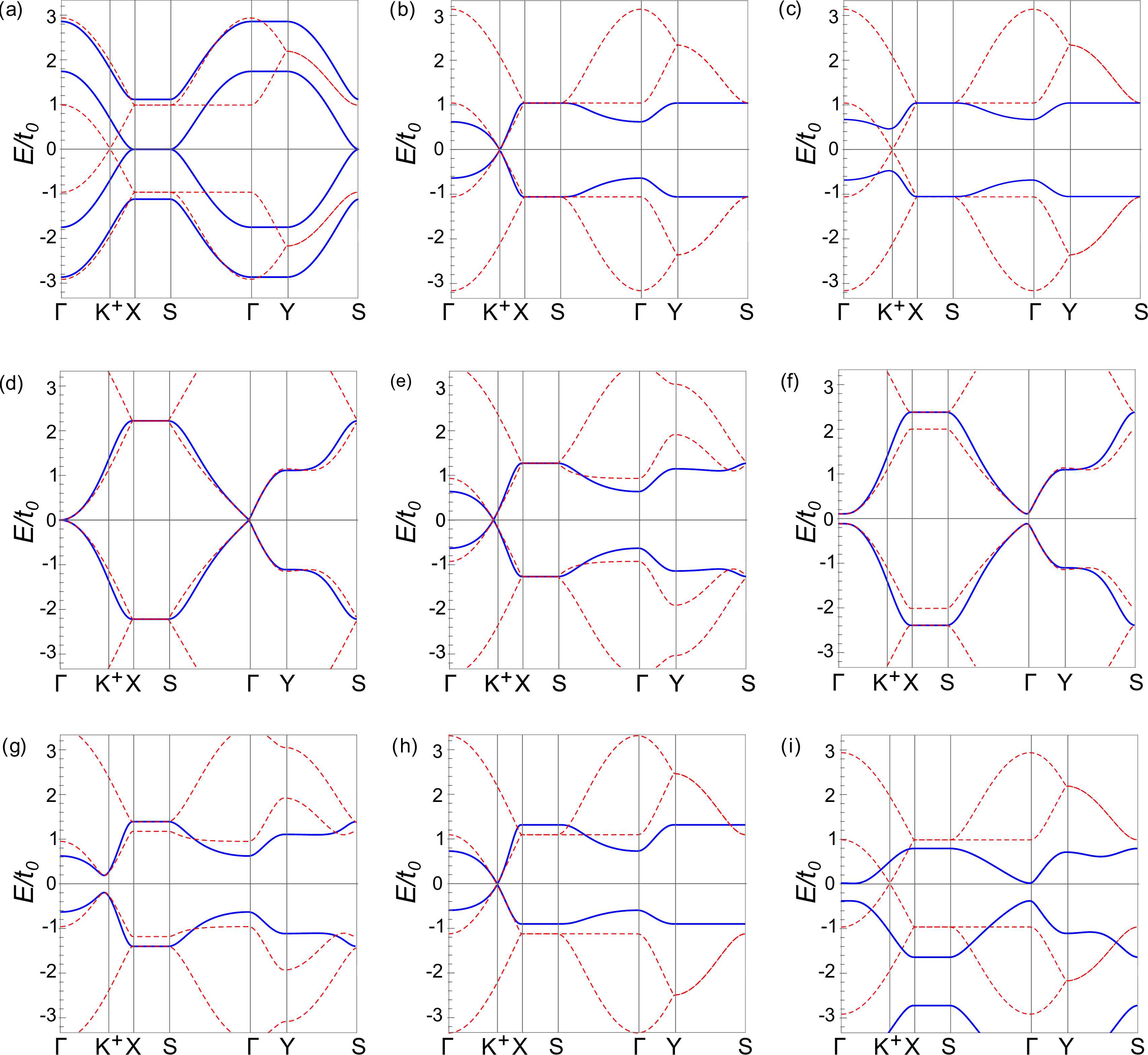}
	\caption{\label{fig_1_SM} (a) Low-energy and dimer bands after setting $r,\,l$ as low-energy and $u,\,d$ as dimer. Low-energy Hamiltonians for (b) fully symmetric case expanded at K$^+$, (c) in the presence of $\mathcal{M}_1(\delta_1)$ at K$^+$, (d) in the presence of $\mathcal{M}_2(\sqrt{3})$ at $\Gamma$, (e) in the presence of $\mathcal{M}_2(\delta_2)$ for generic $k$ point away from K$^+$ determined by the choice of $\delta_2$, (f) in the presence of $\mathcal{M}_1(\delta_1)+\mathcal{M}_2(\sqrt{3})$, at $\Gamma$, (g) in the presence of $\mathcal{M}_1(\delta_1)+\mathcal{M}_2(\delta_2)$, at generic $k$ point away from K$^+$ given by the set of parameters, (h) in the presence of $\mathcal{M}_\text{I}(\delta_\text{I})$ at generic $k$ point given by the set of parameters. This last case is very sensitive to the choice of parameters since it is not able to reproduce properly the flat band. (i) Band structure after adding all three perturbations. Bands are split in energy since all onsite energies are different.} 
\end{figure*}

We obtain a low-energy description of the Hamiltonian~(2) in the main text  by performing first a Taylor expansion around the DCs $\mathbf{k}=\mathbf{K}^++\mathbf{q}$ of all the entries of the matrix, we identify the orbitals contributing to the low-energy features. At this point, we rearrange the Hamiltonian in a low- and high-energy sector $h_\lambda$ and $h_\Delta$, and perform a decimation as introduced in the previous section~\cite{koshino2013electronic}. For this lattice, we have chosen the orbitals placed at lattice sites $r,\,l$ to be the low-energy sector and thus $u,\,d$ are the high-energy one. Figure~\ref{fig_1_SM}(a) shows the spectrum of the low-/high-energy sectors, which is the same if the choice is reversed.
After these two processes, the low-energy Hamiltonian $\tilde{\mathcal{H}}$ for the fully symmetric case reads:
%
%
%%%%%%%%%%%%%
\begin{align}\label{eq:dec_iso}
    \tilde{\mathcal{H}}_0&=v_\text{F} \bm{h}\cdot\bm{\sigma}, \\
    &= v_\text{F}\left[\sqrt{3}(3p_x-p_y)\sigma_x-(p_x-5p_y)\sigma_y\right]. \nonumber
\end{align}
%%%%%%%%%%%%%
%
%
where we have introduced the Fermi velocity defined as $\hbar v_\text{F}=t_0 a_0/4$, $p_\alpha=-\text{i}\hbar\partial_\alpha$ is the momentum operator in the direction $\alpha$, and $\sigma$ are the Pauli matrices. This decimated Hamiltonian reproduces the low-energy physics of the HL properly since we recover the $\pm\pi$ Berry phase that characterizes the cones. The corresponding spectrum is shown in Fig.~\ref{fig_1_SM}(b).
%
%
%%%%%%%%%%%%%%%
%%%%%%%%%%%%%%%
%
%
The low-energy description of the HL with $\left\{\mathcal{G}_{1\alpha}\right\}$ broken corresponds to the following Hamiltonian:
%
%
%%%%%%%%%%%%%%%
\begin{align}\label{eq:dec_g1}
\tilde{\mathcal{H}}_1& =v_\text{F}\bm{h}\cdot\bm{\sigma}\\
& =v_\text{F}\left[f(\bm{p},\delta_1)\sigma_x+g(\bm{p},\delta_1)\sigma_y\right]+t_0\delta_1\sigma_z\nonumber
\end{align}
%%%%%%%%%%%%%%
%
%
where $f(\bm{p},\delta_1)$ and $g(\bm{p},\delta_1)$ are complex functions of their variables which, at $\delta_1
=0$, yield the expressions in~\eqref{eq:dec_iso}. For finite $\delta_1$, the mass term proportional to $\sigma_z$ is finite, thus gaps open at $\mathbf{K}^\pm$ points. The corresponding spectrum is shown in Fig.~\ref{fig_1_SM}(c).
%%%%%%%%
%
%

When breaking the glide $\{\mathcal{G}_{2\alpha}\}$, we obtain a low-energy expansion around $\Gamma$ that recovers the SDC behaviour. It reads:
%
%
%%%%%%%%%%%%%
\begin{subequations}
\begin{align}
\tilde{\mathcal{H}}_2&=v_\text{F}\bm{h}\cdot\bm{\sigma}=-\frac{p_xp_y}{m}\sigma_x+v_\text{F}(p_x-p_y)\sigma_y,\label{eq:dec_g2}\\
E(\mathbf{q})&=\pm\frac{1}{2}\sqrt{16q_y^2-32q_xq_y+q_x^2(16+9q_y^2)}\label{eq:dec_g2_b}.
\end{align}
\end{subequations}
%%%%%%%%%%%%%
%
%
where we have introduced an effective mass defined as $m=t_0/(12v_\text{F}^2)$.
From Eq.~\eqref{eq:dec_g2_b} for small $q$, we find that, by fixing $q_x=q_y$ (\emph{i.e.}~$\Gamma$X direction), we obtain $E\sim q^2$ behavior while, for $q_x=-q_y$ (\emph{i.e.}~$\Gamma$Y direction), we obtain $E\sim |q|$.
At any other $\delta_2$, the series expansion of the decimated Hamiltonian has to be done around the $k$ point given by Eq.~\eqref{eq_G2_kdelta} in the main text to recover the DCs.

Since the Hamiltonian in~\eqref{eq:dec_g2} is written only in terms of two Pauli matrices~\cite{montambaux2018winding}, we can gap the SDC by adding a constant term proportional to $\sigma_z$, which breaks the first set of glides. In the total Hamiltonian, this corresponds to an additional mass term given by the one in the main text, \emph{i.e.}, $\mathcal{M}_1(\delta_1)$.

Figures.~\ref{fig_1_SM}(d) and (e) show the corresponding spectrum for SDC dispersion and general position of the DCs, respectively. Figures.~\ref{fig_1_SM}(f) and (g) show the previous two situations, plus the $\mathcal{M}_1(\delta_1)$ term, where the spectrum is gapped.

Finally, the Hamiltonian for the last choice of onsite energies displays a similar expression as in Eq.~\eqref{eq:dec_g1}.
%
% 
%%%%%%%%%%%%%%%
\begin{figure}[!t]
	\includegraphics[width=\columnwidth]{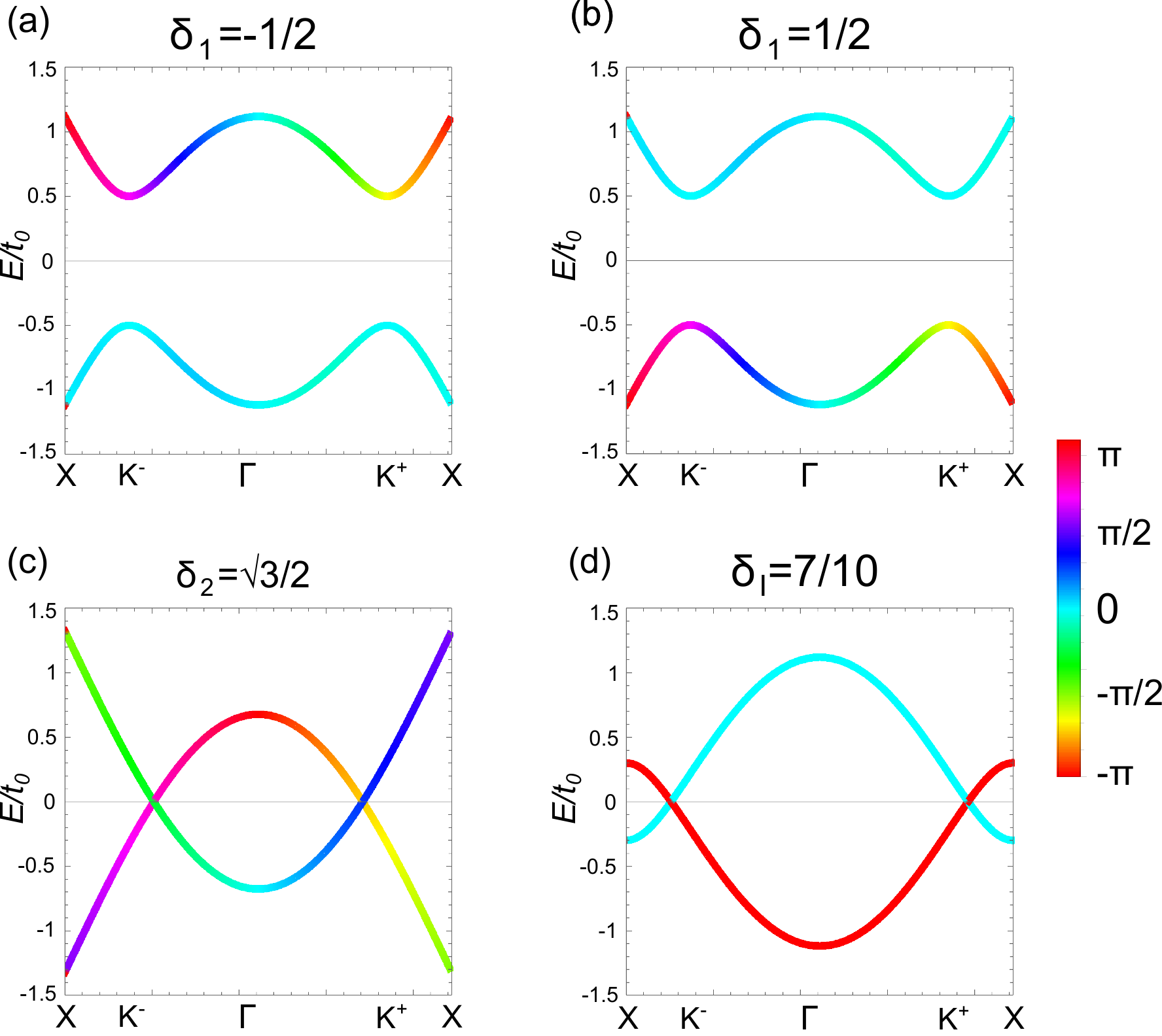}
	\caption{\label{fig_2_SM}(a,b) Symmetry eigenvalues of $\mathcal{G}_{2A}(\mathbf{k})$ in the presence of $\mathcal{M}_1(\delta_1)$ before and after the closing of the gap, (c) in the presence of $\mathcal{M}_2(\delta_2)$ (d) in the presence of $\mathcal{M}_\text{I}(\delta_\text{I})$. In both cases, the crossings are allowed by symmetry eigenvalues, since they are different at the crossing points. The colorbar indicates the phase of the symmetry eigenvalue between $-\pi$ and $\pi$.} 
\end{figure}
%%%%%%%%%%%%%%%
%
%
The corresponding spectrum is shown in Fig.~\ref{fig_1_SM}(h). This spectrum is very sensitive to the choice of parameters. We have chosen a situation where the DC is very close to the fully symmetric case, but, unlike Fig.~\ref{fig_1_SM}(b), the bands are not symmetric in energy.
%%%%%%%%%%%%%%%
\section{Supplementary Note 2: Symmetry eigenvalues}
To study how symmetry eigenvalues evaluate along the bands when the gap opens and closes, we define the $k$-dependent matrix representation of the symmetry operators for all the glides, depending on which orbitals are involved in the operation. We start by studying the components of the spinor $\Psi(\mathbf{r})=(\psi_r(\mathbf{r}),\psi_d(\mathbf{r}),\psi_u(\mathbf{r}),\psi_l(\mathbf{r}))^\text{T}$ in real space after applying the glides:
\begin{align}
    \mathcal{G}_{1A}\Psi(\mathbf{r})&=\begin{pmatrix}
 \psi_u(\mathbf{r}) \\
 \psi_l(\mathbf{r}+\mathbf{a}_2) \\
 \psi_r(\mathbf{r}+\mathbf{a}_1) \\
 \psi_d(\mathbf{r}+\mathbf{a}_1+\mathbf{a}_2) \\
\end{pmatrix}\\
    \mathcal{G}_{1B}\Psi(\mathbf{r})&=\begin{pmatrix}
 \psi_u(\mathbf{r}-\mathbf{a}_2) \\
 \psi_l(\mathbf{r}) \\
 \psi_r(\mathbf{r}+\mathbf{a}_1-\mathbf{a}_2) \\
 \psi_d(\mathbf{r}+\mathbf{a}_1) \\
\end{pmatrix}\\
    \mathcal{G}_{2A}\Psi(\mathbf{r})&=\begin{pmatrix}
 \psi_d(\mathbf{r}+\mathbf{a}_2) \\
 \psi_r(\mathbf{r})\\
 \psi_l(\mathbf{r}+\mathbf{a}_2-\mathbf{a}_1) \\
 \psi_u(\mathbf{r}-\mathbf{a}_1) \\
\end{pmatrix}\\
    \mathcal{G}_{2B}\Psi(\mathbf{r})&=\begin{pmatrix}
 \psi_d(\mathbf{r}+\mathbf{a}_1+\mathbf{a}_2) \\
 \psi_r(\mathbf{r}+\mathbf{a}_1)\\
 \psi_l(\mathbf{r}+\mathbf{a}_2) \\
 \psi_u(\mathbf{r}) \\
\end{pmatrix}
\end{align}
%
%
%%%%%%%%%%%%%
\begin{align}
\mathcal{S}_i=\begin{pmatrix}
    1&0\\
    0& \ee^{\ii k_i}
\end{pmatrix},
\end{align}
%%%%%%%%%%%%%%
%
%
one may define the matrix representation of the glide symmetries as follows:
\begin{subequations}
\begin{align}
\mathcal{G}_{1A}(\mathbf{k})&=
\begin{pmatrix}
 0 & 0 & 1 & 0 \\
 0 & 0 & 0 & \ee^{\ii k_2} \\
 \ee^{\ii k_1} & 0 & 0 & 0 \\
 0 & \ee^{\ii (k_1+k_2)} & 0 & 0 \\
\end{pmatrix}\nonumber\\&=(\mathcal{S}_1\sigma_x)\otimes\mathcal{S}_2,\\
\mathcal{G}_{1B}(\mathbf{k})&=
\begin{pmatrix}
 0 & 0 &  \ee^{-\ii k_2} & 0 \\
 0 & 0 & 0 & 1 \\
 \ee^{\ii (k_1-k_2)} & 0 & 0 & 0 \\
 0 & \ee^{\ii k_1} & 0 & 0 \\
\end{pmatrix}\nonumber\\&=\ee^{-\ii k_2}\mathcal{G}_{1A}(\mathbf{k}),\\
\mathcal{G}_{2A}(\mathbf{k})&=
\begin{pmatrix}
 0 & \ee^{\ii k_2} &  0 & 0 \\
 1 & 0 & 0 & 1 \\
 0 & 0 & 0 & \ee^{\ii (k_2-k_1)} \\
 0 & 0 & \ee^{-\ii k_1} & 0 \\
\end{pmatrix}\nonumber\\&=\mathcal{S}_1^\dagger\otimes(\sigma_x\mathcal{S}_2),\\
\mathcal{G}_{2B}(\mathbf{k})&=
\begin{pmatrix}
 0 & \ee^{\ii (k_1+k_2)} &  0 & 0 \\
 \ee^{\ii k_1} & 0 & 0 & 0 \\
 0 & 0 & 0 & \ee^{\ii k_2} \\
 0 & 0 & 1 & 0 \\
\end{pmatrix}\nonumber\\&=(\sigma_x\mathcal{S}_1\sigma_x)\otimes(\sigma_x\mathcal{S}_2),\\
\mathcal{I}&=
\begin{pmatrix}
 0 & 0 & 0 & 1 \\
 0 & 0 & 1 & 0 \\
 0 & 1 & 0 & 0 \\
 1 & 0 & 0 & 0 \\
\end{pmatrix},
\end{align}
\end{subequations}
%%%%%%%%%%%%%%
%
%
where we have kept the nomenclature used in the main text to name the glides. Now, evaluating the matrix element $\langle u_n(\mathbf{k})|\mathcal{G}_{i,\alpha}|u_n(\mathbf{k})\rangle$ for each $k$ point and band, we can use the phase of such number to obtain the character of such band. The correct glide operator must be used, depending on which one are we breaking. In the case of breaking both glides, we use the inversion operator to study the bands. Figure~\ref{fig_2_SM} shows the symmetry eigenvalues of bands 2\&3 for different perturbations. On the one hand, in the presence of $\mathcal{M}_1(\delta_1)$ after the closing of the gap | Figs.~\ref{fig_2_SM}(a),~\ref{fig_2_SM}(b) | we observe how the characters exchange. On the other hand, in the presence of $\mathcal{M}_2(\delta_2)$ and $\mathcal{M}_\text{I}(\delta_\text{I})$ we study the crossings inside the interval of degeneracy. We should expect that the eigenvalues are different. Otherwise, the bands would go through an avoided crossing. This is precisely what is observed in Fig.\ref{fig_2_SM}(c) and~\ref{fig_2_SM}(d), where at the crossings, the colors are different, and thus the bands go through a proper crossing.

\section{Supplementary Note 3: Combination of all the perturbations}
In this section, we address the effect on the spectrum under the combination of all perturbations: $\mathcal{M}_1,\,\mathcal{M}_2$, $\mathcal{M}_\text{I}$and $q_\text{B}(\mathbf{k})$. This case is the most general that includes all the cases described in the main text. We construct the symmetry operators of the lattice (intra-chain chiral operator $C$, inter-chain chiral operator $C'$, and energy symmetry also known as charge-conjugation-parity, $CP$), and the third Casimir invariant~\cite{Graf_2021}:
\begin{align}
    C&=\tau_z\otimes \sigma_0,\\
    C'&=\tau_z\otimes(e^{ik_1} \sigma_++ e^{-ik_2} \sigma_-),\\
    CP&= K \tau_y \sigma_x,\\
    \mathcal{C}_3&=\Tr(h(\mathbf{k})^3),
\end{align}
where $\sigma_\pm$ are symmetric and antisymmetric linear combinations of $\sigma_{x,y}$, and $K$ is the complex conjugation operator. The chiral symmetry operators plus the energy symmetry operator are helpful to understand when the energy spectrum is energy symmetric. However, the third Casimir invariant is a more straightforward approach for detecting energy symmetric spectra since it can be non-zero even if trace of the Hamiltonian is. Four band systems are energy symmetric if the third Casimir invariant is zero. In the following we summarize the combinations of all perturbations that respect or break the energy symmetry:
\\
\begin{enumerate}
    \item $\mathcal{M}_1+\mathcal{M}_2+\mathcal{M}_\text{I}\to$ No energy symmetry.
    \item $q_\text{B}(\mathbf{k})+\mathcal{M}_1\to$ Energy symmetry,%
    \item $q_\text{B}(\mathbf{k})+\mathcal{M}_2\to$ Energy symmetry,
    \item $q_\text{B}(\mathbf{k})+\mathcal{M}_\text{I}\to$ No energy symmetry,
    \item $q_\text{B}(\mathbf{k})+\mathcal{M}_1+\mathcal{M}_2\to$ Energy symmetry,
    \item $q_\text{B}(\mathbf{k})+\mathcal{M}_1+\mathcal{M}_\text{I}\to$ No energy symmetry,
    \item $q_\text{B}(\mathbf{k})+\mathcal{M}_2+\mathcal{M}_\text{I}$ $\to$ No energy symmetry,
    \item $q_\text{B}(\mathbf{k})+\mathcal{M}_1+\mathcal{M}_2+\mathcal{M}_\text{I}\to$ No energy symmetry.
\end{enumerate}

Regarding the first case with all the onsite energies, we obtain $\Tr(h(\delta_1,\delta_2,\delta_\text{I},\mathbf{k}))=0$ for all values of $(\delta_1,\delta_2,\delta_\text{I})$, so one may expect to obtain energy symmetry. However:
\begin{equation}
\mathcal{C}_3(\delta_1,\delta_2,\delta_\text{I})=\Tr(h(\delta_1,\delta_2,\delta_\text{I},\mathbf{k})^3)=24 \delta_1\delta_2\delta_\text{I},\label{eq_trH3allonsites}
\end{equation}
which is different from zero for generic values of the parameters $(\delta_1,\delta_2,\delta_\text{I})$ and thus energy symmetry is not recovered. Figure~\ref{fig_1_SM}(i) shows the band structure for $\mathcal{M}_\text{T}=\mathcal{M}_1(0.5)+\mathcal{M}_2(2)+\mathcal{M}_\text{I}(1)$, where the highest band does not appear since it is very high in energy. The Casimir invariant for the most general case with all onsites plus dimerization (item \#8) yields:
\begin{equation}
\mathcal{C}_3(\delta_1,\delta_2,\delta_\text{I},\delta_{\text{D}})=-24(\delta_{\text{D}}- \delta_1\delta_2)\delta_\text{I},
\end{equation}
which is non-zero for generic values and returns \eqref{eq_trH3allonsites} at $\delta_\text{D}=0$.

\end{document}